# MICROWAVE BACKGROUND, BACKGROUNDS HIERARCHY - POLYPHONY OF UNIVERSE

**Chechelnitsky A.M.** E'mail: ach@thsun1.jinr.ru

## ABSTRACT

We discuss the structure, physical nature, dynamic genesis of clearly detected diffuse backgrounds (radiations) in the Universe. From the point of view of Wave Universe concept it is shown, that observed Backgrounds Hierarchy (Polyphony of Universe) is close connected and causely determined by the *sound velocity* Hierarchy of the physical medium, *cosmic (polycomponent)* plasma. The well-known Microwave Background (Cosmic Background Radiation - CBR) corresponds to the $F^{[3]}$ Background, that is characterized by the calculated sound velocity $C_*^{[3]}$ =11.483 km$\cdot$s$^{-1}$, by the calculated temperature $T_*^{[3]}$=2.9°K and by wavelength of the radiation intensity maximum $\lambda_*^{[3]}$=1 mm.

## HIERARCHY OF SOUND VELOCITIES AND TEMPERATURES

In the frame of Wave Universe concept, megawave and Shell structure of astronomical systems, considering as *wave dynamic systems* (WDS) [Chechelnitsky, 1980-2001], in the polycomponent cosmic plasma there is the hierarchy of velocities of disturbances propagations (megawaves) - of the *sound velocities*

$$C_*^{[s]} = (1/\chi^{s-1}) C_*^{[1]}, \quad s = ..., -2, -1, 0, +1, +2, ...$$

where

$C_*^{[1]}$ = 154.3864 km$\cdot$s$^{-1}$ is the *calculated* value of *sound velocity* of wave dynamic system (WDS) in the $G^{[1]}$ Shell, that was made valid by observations in the Solar System,

$\chi$ = 3.66(6) - the *Fundamental parameter of hierarchy – Chechelnitsky Number* [Chechelnitsky, 1980 - 1986],

s - the countable parameter of $G^{[s]}$ Shells.

To each physically distinguished (elite) velocity $C_*^{[s]}$ of this series the electron temperature $T_*^{[s]}$ may correspond, the value of T may be calculated by virtue of standard common relation

$$T = (1/3\,k) m_e v^2, \quad T = (v/6.743)^2\ [K°], \quad v[km\cdot s^{-1}],$$
$$v = C_*^{[s]},$$

where $m_e$ - mass of electron, k - Boltzman constant, and thus the hierarchy series of *electron temperatures*

$$T_*^{[s]} = (m_e/3k)\cdot(C_*^{[s]})^2 = T_*^{[1]}/\chi^{2(s-1)}, \quad s = ...,-2,-1,0,+1,+2,...$$
$$T_*^{[1]} = (m_e/3k)\cdot(C_*^{[1]})^2 = 524.21\ K°,$$

may be obtained.

The Table contains data about the hierarchy of physical parameters (of $G^{[s]}$ Shells)

$$\lambda_* = b/T_*, \quad b = 2.89779\cdot 10^{-3} [m\cdot K°] \quad \text{(from Wien's Law)},$$
$$\nu_* = c/\lambda_*, \quad \text{(c - light velocity)},$$
$$T_*[ev] = T_*[K°]/11604.5$$

of astronomical systems, that characterize the structure of cosmic medium - polycomponent cosmic plasma.

## HIERARCHY OF ELITE VELOCITIES AND TEMPERATURES

In the general case *elite (dominant) velocities* [Chechelnitsky, 1980-2001] are given the by formula

$$v_N^{[s]} = C_*^{[s]} (2\pi)^{1/2}/N, \quad s = ...,-2,-1,0,1,2,...$$
$$C_*^{[s]} = (1/\chi^{s-1})\cdot C_*^{[s]}.$$

Here N - (Mega)Quantum numbers of *elite states*,

a) Close to

$N_{dom}$= 8; 11; 13; (15.5)16; (19,5); (21,5) 22,5 -

for the *strong elite (dominant) states* (orbits);

b) Close to

N - Integer, Semi-Integer -

for the *week elite (recessive)* states (orbits).

In the wave structure of the Solar System for planetary orbits of Mercury (ME), Venus (V), Earch (E), Mars (MA), we have, in particular, $N = (2\pi a/a_*)^{1/2}$ (a - semi-major axes of planetary orbits, $a_*^{[1]} = 8R_\odot$ - semi-major axis of $TR_*^{[1]}$ - Transsphere, $R_\odot$ - radius of Sun) [Chechelnitsky, 1986]

N = 8.083; 11.050; 12.993; 16.038, close to *integer*

N = 8; 11; 13; 16.

Take into account Ceres (CE) orbit and transponated in $G^{[1]}$ (from $G^{[2]}$) planetary orbits of Uranus - (U), Neptune - (NE), Pluto - (P), it can be received the general representation for observational dominant N

| | $TR_*$ | ME | TR | V | E | (U) | MA | (NE) | CE | (P) |
|---|---|---|---|---|---|---|---|---|---|---|
| N= | $(2\pi)^{1/2}$=2.5066 | 8.083 | $(2\pi)^{1/2}\chi$=9.191 | 11.050 | 12.993 | 15.512 | 16.038 | 19.431 | 21.614 | 22.235 |

Hence in general case we get the result for $T_N^{[s]}$ electron temperatures

$$T_N^{[s]} = T_*^{[s]}(2\pi/N^2),$$

For $N = N_* = (2\pi)^{1/2}$=2.5066 we get the above result for sound velocities and result for critical temperature

$$T_*^{[s]} = T_*^{[1]}/\chi^{2(s-1)}, \quad s = ...,-2,-1,0,+1,+2,...$$

## MICROWAVE BACKGROUND

It is interesting, that in the hierarchy series of $T_*^{[s]}$ electronic temperatures - in case s=3 it may be exposed the temperature $T_*^{[3]}$ = 2.90° K which corresponds to $C_*^{[3]}$=11.483 km·s$^{-1}$ sound velocity (in $G^{[3]}$ Shell), wavelength of the radiation intensity maximum $\lambda_*^{[3]}$=1mm and coinsides with observed temperature $T_*$=2.9° K of cosmic background radiation - microwave background (maximum intensity of radiation - at the wavelength $\lambda_{max}\approx$1 mm) - see Table, [Longair, 1984, p.202-203, Fig. 15.13], [Woody and Richards, 1981; Longair, 1983, p. 316, Fig. 15.2].

The used *calculated* of the $T_*^{[3]}$ temperature of microwave background assume some refinement in connection with the possible more precise definition of the $C_*^{[3]}$ sound velocity of cosmic plasma in $G^{[3]}$ Shell. That sound velocity (as connected with it $C_*^{[1]}$ and $C_*^{[s]}$ sound velocities) represents the fundamental characteristic of the Wave Universe and is subject to very careful definition and correction by special purposeful observations and experiments.

## BACKGROUNDS HIERARCHY - POLYPHONY OF UNIVERSE

In the $C_*^{[s]}$ sound velocities hierarchy the $C_*^{[3]}$ = 11.483 km·s$^{-1}$ velocity, generally speaking, is not special, preferable in relation to any $C_*^{[s]}$ velocities. Therefore, it may be expected the existence not only corresponding $T_*^{[3]}$=2.9°K temperature, but also another $T_*^{[s]}$ temperatures (see Table) - the existence of Universe Polyphony.

It should be interesting to verify if correspond to its in observations and experiments another (high temperature, high energetic) diffuse backgrounds $F^{[s]}$ possessing (as the $F^{[s]}$ microwave background) maximum of radiation intencity at wavelengths $\lambda_{max}=\lambda_*^{[s]}$ (see Table).

The positive result of purposeful observations and experiments can make, in particular, more clear, observable fundamental physical property of both microwave background and another (in particular, X-Rays, Gamma-Rays, etc.) backgrounds (phones) - its (local) isotropy, gomogeneity, informly filling of the sky sphere, its diffuse character. This is consequence of the fundamental dynamical property - isotropy of cosmic medium in each $G^{[s]}$ Shell (propagation in it is isotropy, sound velocities are equal for any directions of propagation).

## OBSERVATIONS OF ANOTHER DIFFUSE BACKGROUNDS

In the present time existing observational data demonstrate existence:

Besides of $F^{[3]}$ Microwave Background, still at least,

# Diffuse Radio Background $F^{[7]}$("plateau " in the 10 MHz region and "fall" at more high frequencies - see [Fig.1.6, p.40 in Galactical...,1976]) and also

# Diffuse $F^{[6]}$ Background ($\nu_*^{[6]}$=123.46 MHz) and $F^{[5]}$ Background ($\nu_*^{[5]}$=1659.9 MHz);

# Diffuse Infra Red (IR) Background in the $\lambda \approx 5$ μm region [Matsumoto et al., 1984], that is associated with $F^{[1]}$ IR-Background ($\lambda_*^{[1]}$ = 5.527 μm);

# Soft X-Rays Background in region 0.10 - 0.28 kev [Marchall et al., 1984], that is associated with $F^{[-2]}$ Background ($T_*^{[-2]}$ = 0.109 kev);

# Diffuse X-Rays Background in region 1÷1.5 kev [Rocchia, 1983] that is associated with $F^{[-3]}$ Background ($T_*^{[-3]}$ = 1.475 kev);

# Diffuse Gamma Background in region 0.28÷4.35 Mev [Jayanthi, 1983], that is associated with $F^{[-5]}$ ($T_*^{[-5]}$ = 0,266 Mev) and $F^{[-6]}$ ($T_*^{[-6]}$ = 3.586 Mev) Backgrounds.

## MANY YEARS LATER

More then decade go away from the original date of promulgation and discussion of Universe Poliphony concept [Chechelnitsky, 1986 a,b,c]. Additonal received at that time observational information, correspondence observations to the theory prediction only intensify the positions of above discussed representation. As before in the center of view of astrophysics and cosmology stay the problem of physical nature, fundamental genesis of microwave background (CBR) and another observed diffuse background radiations of Universe. And as before many aspects of standard theory representations remain extremely vulnerable for the criticism. We point only some of them.

## DYNAMICAL STATUS OF MICROWAVE BACKGROUND

From the view of Wave Universe concept hardly it has not a sense to attribute *to only* microwave background (Cosmic Background Radiation - CBR) any special (distinguished from another backgrounds) status.

Frequentely attached to it the term - *Relict* - attempts to underline its special status. Evidently, this term has conceptual load, connected with special representation about its genesis in the frame of Hot Universe and Big Bang concept.

But why the cosmology may consider only the one - microwave - background as *remainder-relict* of Big Bang, but whole set of backgrounds (close near it displaced) - *is not*?

In the frame of carring analysis and common sense it is represented as the more argumented principle - either *all backgrounds are relict* (?), or - *not a single* (!).

Wave Universe concept insists on *actual (today's)* (but not on inherited from event outstanding by billion years ago) nature of the microwave background, more exactly,
of microwave backgrounds. Becouse any wave dynamic system (WDS) of Universe, such local astronomical system as the Solar system, our Galaxy, Local Superclaster (Virgo) is characterized *by own* $G^{[3]}$ Shells and corresponding $F^{[3]}$ backgrounds.

And then, observing the Heaven from the Earth, we must detect not entirely homogeneous, isotropic background (and not a single), but some superposition, covering of microwave backgrounds of these (Local) astonomical systems of more and more growing scales (inside of those is enclosed an Observer).

And then it becomes comprehend in more degree the phenomenon of observing with high resolution in the last time *pecularities* of the cosmic background radiation (CBR) (microwave background).

It may be wait, for instance, that with multiple precision of observational astronomy the region of *own microwave background* of Our Galaxy (in the form of distinguished pecularity) will be remind (in the Galaxy system of coordinates) some *lense* (similar to radiogalo), which contains the more compact part of Galaxy.

And indeed, at [Bennet et al., 1996; Lasenby and Hancock, 1997, p. 220, Fig.3] it can see this forseeing in the frame of Wave Universe concept the *local microwave background lense* of our Galaxy - as *nonhomogeneity* in measurment of the temperature of cosmic background radiation (CBR) on COBE satellite.

Observed in infrared rays giant shell (corona) of NGC 4565 galaxy [Spinrad et al., 1978; Marochnik and Suchkov, 1984, Fig. 61, Fig.1] is the bright illustration of existence of *another local backgrounds* and connected with its *local Shells* in far astronomical objects. Evidently, this giant corona connect with $F^{[1]} \div F^{[3]}$ local IR Backgrounds and with corresponding $G^{[1]} \div G^{[3]}$ Shells of this galaxy.

**Table** **BACKGROUNDS HIERARCHY – POLYPHONY OF UNIVERSE**

| $G^{[s]}$ Shell Index S | Critical Redshift $Z_*$ | Critical Temperature $T_*$ [$K^o$] | Critical Temperature $T_*$ [ev] | Critical Wave Length of Max Radiation $\lambda_*$ [m] | Critical Frequence $\nu_*$ [Hz] | Diffuse Backgrounds $F^{[s]}$ | |
|---|---|---|---|---|---|---|---|
| -10 | 0.688 E+06 | 0.135 E+16 | 0.117 E+12 | 0.213 E –17 | 0.140 E+27 | $F^{[-10]}$ | Cosmic |
| -9 | 0.512 E+05 | 0.101 E+15 | 0.871 E+10 | 0.286 E -16 | 0.104 E+26 | $F^{[-9]}$ | Rays |
| -8 | 0.381 E+04 | 0.752 E+13 | 0.648 E+09 | 0.385 E -15 | 0.778 E+24 | $F^{[-8]}$ | |
| -7 | 0.283 E+03 | 0.559 E+12 | 0.482 E+08 | 0.517 E -14 | 0.578 E+23 | $F^{[-7]}$ | |
| -6 | 0.211 E+02 | 0.416 E+11 | 0.358 E+07 | 0.696 E -13 | 0.430 E+22 | $F^{[-6]}$ | Gamma |
| -5 | 0.157 E+01 | 0.309 E+10 | 0.266 E+06 | 0.936 E -12 | 0.320 E+21 | $F^{[-5]}$ $10^{-11}$ m | Rays |
| -4 | 0.116 E+00 | 0.230 E+09 | 0.198 E+05 | 0125 E -10 | 0.238 E+20 | $F^{[-4]}$ | |
| -3 | 0.866 E -02 | 0.171 E+08 | 0.147 E+04 | 0.169 E -09 | 0.177 E+19 | $F^{[-3]}$ $10^{-9}$ m | X-Rays |
| -2 | 0.644 E -03 | 0.127 E+07 | 0.109 E+03 | 0.227 E -08 | 0.131 E+18 | $F^{[-2]}$ | Soft X-Rays |
| -1 | 0.479 E -04 | 0.947 E+05 | 0.816 E+01 | 0.305 E -07 | 0.980 E+16 | $F^{[-1]}$ $10^{-7}$ m | UV |
| 0 | 0.357 E -05 | 0.704 E+04 | 0.607 E+00 | 0.411 E -06 | 0.729 E+15 | $F^{[0]}$ $10^{-6}$ m | Opt |
| 1 | 0.265 E -06 | 0.524 E+03 | 0.451 E -01 | 0.552 E -05 | 0.542 E+14 | $F^{[1]}$ | Infra - Red (IR) |
| 2 | 0.197 E -07 | 0.389 E+02 | 0.336 E -02 | 0.743 E -04 | 0.403 E+13 | $F^{[2]}$ | |
| **3** | **0.147 E -08** | **0.290 E+01** | **0.249 E-03** | **0.999 E-03** | **0.300 E+12** | $\mathbf{F^{[3]}}$ **$10^{-3}$ m** | **Microwave (CBR)** |
| 4 | 0.109 E -09 | 0.215 E+00 | 0.185 E -04 | 0.134 E -01 | 0.223 E+11 | $F^{[4]}$ | |
| 5 | 0.812 E -11 | 0.160 E -01 | 0.138 E -05 | 0.180 E+00 | 0.165 E+10 | $F^{[5]}$ | Radio ( R ) |
| 6 | 0.604 E -12 | 0.119 E -02 | 0.102 E -06 | 0.242 E+01 | 0.123 E+09 | $F^{[6]}$ 10 m | |
| 7 | 0.449 E -13 | 0.887 E -04 | 0.764 E -08 | 0.326 E+02 | 0.918 E+07 | $F^{[7]}$ | R |
| 8 | 0.334 E -14 | 0.660 E-05 | 0.568 E -09 | 0.438 E+03 | 0.683 E+06 | $F^{[8]}$ $10^{3}$ m | |
| 9 | | 0.490 E -06 | | 0.59 E+04 | | $F^{[9]}$ | R |